\documentclass[aps, twocolumn, prb, 10pt, floatfix]{revtex4-1}

\usepackage{amssymb}
\usepackage{amsmath}
\usepackage{graphicx}
\usepackage{bm}
\usepackage{mathtools}
\usepackage{tikz}
\usetikzlibrary{arrows,positioning} 
\usepackage{wrapfig}
\usepackage{hyperref}
\usepackage{longtable}
\usepackage{CJKutf8}
\usepackage{cancel}
\usepackage{todonotes}
\usepackage{braket}
\usepackage{multirow}

\begin{document}
\title{Electrons, holes, and spin in the IV-VI monolayer `four-six-enes'}
\author{Ian Appelbaum}
\email{appelbaum@physics.umd.edu}
\author{Pengke Li (\begin{CJK*}{UTF8}{gbsn}李鹏科\end{CJK*})}
\email{pengke@umd.edu}
\affiliation{Department of Physics and Center for Nanophysics and Advanced Materials, University of Maryland, College Park, Maryland 20742, USA}

\begin{abstract}
Bandedge states in the indirect-gap group-IV metal monochalcogenide monolayers (`four-six-enes' such as SnS, GeTe, etc.) inherit the properties of nearby reciprocal space points of high symmetry at the Brillouin zone edge. We employ group theory and the method of invariants to capture these essential symmetries in effective Hamiltonians including spin-orbit coupling, and use perturbation theory to shed light on the nature of the bandedge states. In particular, we show how the structure of derived wavefunctions leads to specific dominant momentum and spin scattering mechanisms for both valence holes and conduction electrons, we analyze the direct optical transitions across the bandgap, and expose the interactions responsible for subtle features of the local dispersion relations.
\end{abstract}

\maketitle

\section{Introduction}

Broken lattice symmetry nearly always has substantial consequences for the electronic structure of materials\cite{Dresselhaus_B08, BC_B10, Yu_B10}. One well-known example is Dresselhaus spin splitting \cite{Dresselhaus_PR55} upon the breaking of inversion symmetry in any time-reversal invariant (i.e. non-magnetic) system; this is fundamental to the differences between bandstructure in the case of diamond and zincblende lattices. Another such phenomenon is the valley splitting of otherwise equivalent band extrema in the presence of uni- or bi-axial strain that breaks a discrete rotational symmetry in any multivalley material, e.g. the group-IV elemental semiconductors Si or Ge \cite{Herring_PR56}. Even a subtle atomic distortion, such as in the transition from cubic to tetragonal phase in many perovskites \cite{Mattheiss_PRB72}, has dramatic repercussions on band dispersion beyond the obvious generation of multiferroic moments.

Within the realm of two-dimensional semiconductors, phosphorene (see Ref.~[\onlinecite{Li_PRB14}] and references therein) provides an elemental touchstone material on which to explore the role of symmetry breaking. Removing the equivalence of all four group-V atoms within the unit cell by replacing them with equal numbers of group-IV and -VI atoms to retain the complete $sp^3$ covalency has many immediate consequences: Dresselhaus spin splitting will emerge due to broken space inversion symmetry, and together with a persistent uniaxial deviation from orthorhombic configuration, the more ionic bonding between atoms with different valence results in a bulk dipolar electric field\cite{Fei_APL15,Gomes_PRB15b}. Also, the relevant band-edge states are no longer at the Brillouin zone center, so that remaining structural anisotropy gives rise to valley splitting. 

The resulting materials formed from \{Ge, Sn\} and \{S, Se, Te\} are the so-called `group-IV metal monochalcogenide' monolayers. More succinctly named `four-six-enes' (adapted from `graphene'), they have been studied recently using density functional theory (DFT) both with and without spin-orbit interaction\cite{Gomes_PRB15, Hu_APL15, Kamal_PRB16}. While this well-developed method provides valuable information on band ordering and overall dispersion, the essential (and often simple) physical reasons for apparent bandstructure characteristics peculiar to the four-six-enes remain obscured behind the lid of DFT's black box, where fundamental symmetries in the underlying Schr\"odinger equation are buried.

In the present paper, we use the mathematical language of symmetry (group theory) to provide the answers to intriguing questions posed, and yet unexplained, by DFT band calculations of four-six-ene structures. Although the concepts of symmetry and corresponding irreducible representations (IRs) were previously mentioned in the context of bandstructure in these materials \cite{Rodin_PRB16}, they were mistakenly used by attributing the properties of salient gap-edge conduction band (CB) and valence band (VB) states nearby the Brillouin zone (BZ)-edge to the point group, which is only strictly relevant to the Brillouin zone-center $\Gamma$-point. In contrast, we utilize the full theory of the space group and method of invariants for nearby $k$-points of high symmetry to construct concise effective Hamiltonians which not only yield correct CB and VB band dispersion, but -- more importantly -- capture the spin-dependent nature of electronic states. In addition to elucidating clear symmetry-borne band interactions inducing the dispersion characteristics apparent in DFT results, we analyze the wavefunctions to address selection rules for interband optical transitions and relaxation of both momentum and spin. 

The outcome of this analysis will reveal many enticing qualities of four-six-enes otherwise not evident solely from dispersion along the axes of high symmetry. In the spin-split lowest conduction band valleys, we find that the dominant scattering mechanism in many 2D materials (carrier interaction with flexural phonons that have quadratic dispersion\cite{Mariani_PRL08, Song_PRL13}) is relevant here only to high-order scattering processes, and electron mobility is limited by the presence of acoustic phonons. When considering spin relaxation, we explain that Elliott-Yafet spin-flip mechanisms are suppressed in this band by a fortuitous near-cancellation of contributions to spin-mixing amplitudes in a prototypical four-six-ene.
At the spin-degenerate valence band maximum, we find that armchair-polarized acoustic phonons limit the hole mobility whereas flexural phonons are a dominant cause of spin-flip transitions. Away from the high-symmetry axis, a linear Dresselhaus field\cite{Rashba_FTT59} (similar to that found in zincblende [110] quantum wells) provides an environment capable of supporting an in-plane persistent spin helix for polarized holes. 
Direct optical transitions across the fundamental gap between these bands (but not between absolute extrema) induced by linearly polarized light are controlled by momentum matrix elements, whose relative magnitude (when nonzero) can be easily explained using symmetry arguments.

The organization of this paper is as follows. In Sec. \ref{sec:DFT} we present the DFT-calculated bandstructure including spin-orbit interaction of a prototypical and naturally-occurring example of a four-six-ene, tin (II) sulfide (SnS). We identify several intriguing properties evident in the dispersion close to the CB and VB extrema states, and pose questions on their origins common to all four-six-enes. In Sec. \ref{sec:symmetry}, we discuss the geometric symmetry of the real-space lattice and describe the point group. We extend this to space group of the two relevant high-symmetry $k$-points at the edge of the BZ that are in close proximity to the band extrema, and derive the effective Hamiltonians in Secs. \ref{sec:Ypoint} and \ref{sec:Zpoint}. Selection rules for phonon- and photon-induced transitions specific to each case are discussed within the corresponding section.  Finally, in Sec. \ref{sec:conclusion} we comment on the suitability of this material's electronic structure for potential applications and observation of unusual phenomena, such as optical creation of a degenerate exciton condensate.

\section{Bandstructure interrogation \label{sec:DFT}}

\begin{figure}
\includegraphics[width=8cm]{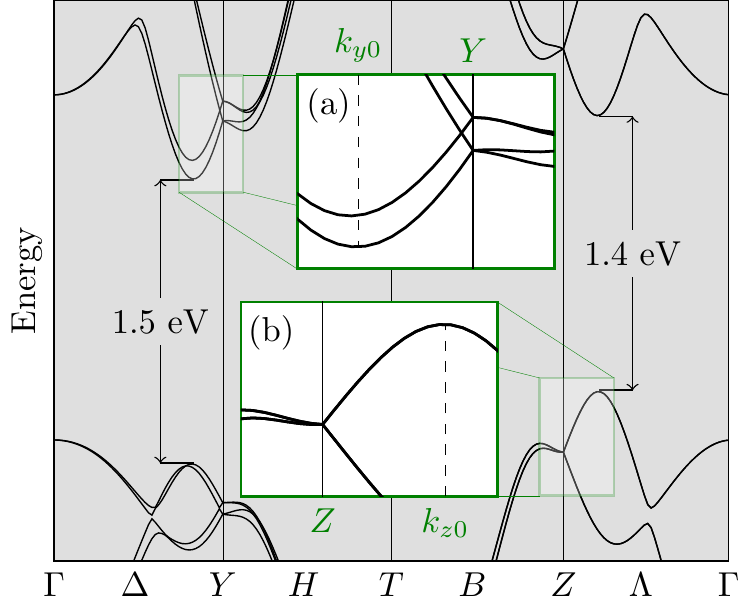}
\caption{(Color online) Bandstructure of monolayer SnS `four-six-ene' including spin-orbit interaction, calculated by the {\sc Quantum ESPRESSO} package\cite{QE-2009} using PBEsol functional with Projector Augmented-Wave (PAW) fully relativistic pseudopotentials, 40 Ry planewave cutoff, 180 Ry charge density cutoff, and $8\times 8\times 2$ Brillouin zone sampling grid.  The relaxed lattice constants are $a_z =$~4.35~\AA\, and $a_y = 4.02$ \AA, and the distance between isolated layers is $3.9a_z$. The symmetry analysis throughout this paper is robust to variations in these numerical details. Insets (a) and (b) zoom in on the details of bandstructure close to the highly symmetric $Y$- and $Z$-points and their associated conduction and valence valleys, respectively. \label{fig:DFT}}
\end{figure}

Using \textit{ab initio} density-functional theory (DFT) including spin-orbit interaction with the {\sc Quantum ESPRESSO} package,\cite{QE-2009} we have calculated the bandstructure of tin (II) sulfide, SnS, and obtain a dispersion nominally identical to other reports\cite{Gomes_PRB15, Hu_APL15, Kamal_PRB16}. The results along high symmetry axes on the rectangular irreducible Brillouin zone boundary [given in Fig. \ref{fig:lattice}(b)] are shown in Fig. \ref{fig:DFT}, where the regions around band extrema close to the $Y$- and $Z$-points are emphasized in insets (a) and (b), respectively. These high-symmetry point labels are chosen to be consistent with Ref. \onlinecite{BC_B10}, Fig. 3.5. 

Because these extrema are situated at points that are not time-reversal (TR) invariant, four-six-enes are multivalley semiconductors with two equivalent extrema valleys in each band, $\approx$80\% from the BZ center $\Gamma$-point. Valley splitting causes the valence band maximum (VBM) and conduction band minimum (CBM) to be on different axes, so the fundamental gap is indirect. 

Looking closer, we first notice the difference in band degeneracy at these points of high symmetry. Clearly these states at $Y$ are doubly degenerate, whereas those at $Z$ result from the culmination of two orbital bands, each themselves spin degenerate. We also see other details, such as an accidental crossing of oppositely-dispersing orbital bands close to $Y$ and an obvious spin-orbit-induced splitting of the conduction band minimum. No such splitting occurs at the valence band maximum away from $Z$.

What symmetries dictate these bandstructure features? What are the wavefunctions at band extrema? What are the selection rules for optical transitions between bands? What kind of phonon scattering is allowed, and what effect does it have on the charge mobility and spin relaxation? All of these questions can be answered by analyzing the symmetry of the system to determine the behavior of states at the high-symmetry points $Y$ and $Z$, and extending it to the extrema along the $\Delta$ and $\Lambda$ axes. This strategy requires the appropriate use of the method of invariants to construct effective Hamiltonians around these high symmetry points, trivial analytic diagonalization of 2$\times$2 matrices, and lowest-order perturbation theory to expand to finite $k$ and include the effects of spin-orbit interaction. In the next section, we discuss the geometric configuration of the atomic lattice and describe the rotation, reflection, and partial translation operations that leave the lattice invariant, as our first step in developing the group theory for four-six-ene electronic structure.

\section{Spatial symmetry\label{sec:symmetry}}
\subsection{Lattice basics} 

\begin{figure}
\includegraphics{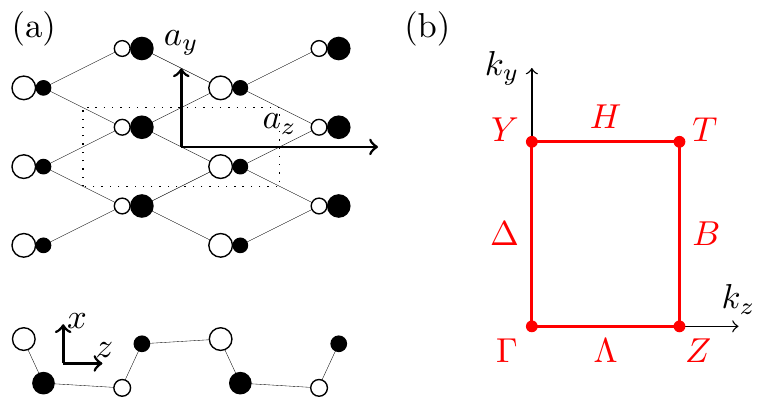}
\caption{(Color online) (a) Real-space lattice, where white and black circles represent different group-IV and -VI atoms and size represents out-of-page displacement. Plan-view is shown at top of panel, whereas bottom shows the side as viewed along the zig-zag ($y$) direction.; (b) Reciprocal-space lattice showing irreducible Brillouin zone. Our coordinate system follows the labeling convention established in Ref.~[\onlinecite{BC_B10}].\label{fig:lattice}}
\end{figure}
As shown in Fig. \ref{fig:lattice}(a), the phosphorene-related four-six-enes have an orthorhombic `distorted rocksalt' lattice that results from `puckering' a boron nitride-like honeycomb lattice along the zigzag $y$-direction. This uniaxial warping results from the increased average atomic valence of five and consequent tetragonal $sp^3$ coordination, in contrast to boron nitrides's average valence of four and planar $sp^2$ bonding. The rectangular unit cell is then shorter in the $y$-direction than in the orthogonal in-plane $z$-direction, causing a rectangular BZ as shown in Fig. \ref{fig:lattice}(b) and an obvious anisotropy in the bandstructure including valley splitting. Importantly, a polar distortion causes out-of-plane bonds to tilt along the $z$-direction, which is responsible for breaking inversion symmetry otherwise present in the undistorted rocksalt lattice.

\subsection{Point-group symmetry} 

The four-six-ene lattice thus has a remarkably reduced symmetry. Other than translations by integral numbers of unit vectors, there are only four symmetry operators that leave the atomic configuration invariant: $E$ (the identity operator); $\tau C_{2z}$ [rotation by $180^{\circ}$ around the $z$ axis then translation of $\tau=(\frac{a_y}{2},\frac{a_z}{2})$]; $\tau R_x$ (reflection with respect to the $zy$ plane, then  translation by $\tau$); and $R_y$ (reflection with respect to the $xz$ plane). We make note in particular of the glide operations involving $\tau$, which make this group non-symmorphic. 

In comparison to $D_{2h}^{18}$ phosphorene, the symmetry group for four-six-enes is isomorphic to $C_{2v}^7$ in Sch\"onflies notation, or \#31 ($Pmn2_1$) in International/Hermann-Mauguin notation. Each of the aforementioned four group elements forms its own class $\mathbb{C}_{1-4}$, respectively, due to the abelian nature of the $C_{2v}$ point group.

If only states near the BZ center $\Gamma$-point were relevant to our discussion (as in phosphorene\cite{Li_PRB14}), we could present the irreducible representations and their characters, basis functions and invariant components from the $\bm k\cdot \hat {\bm p}$ Hamiltonian corresponding to this closed set of symmetry operators. Then we would classify the wavefunctions and their symmetry-allowed interactions giving rise to band dispersion. However, the region of reciprocal space salient to the conduction and valence band edge states in four-six-enes is clearly at far away from $k=0$, as shown in Fig. \ref{fig:DFT}. This fact makes the point group an incomplete catalog of symmetries for physical wavefunctions.

Although we have a tabulation of the symmetry operations for the Bloch envelope function that fully captures the transformation properties of wavefunctions at the $\Gamma$-point, the multiplicative planewave component $e^{i\bm {k} \cdot \bm {r}}$ also contributes to symmetry properties at $k\neq~\!\!0$. As mentioned in a previous symmetry analysis of phosphorene\cite{Voon_NJP15}, the nonsymmorphic nature of the lattice results in band merging and double degeneracy at the Brillouin zone edge. Specifically, even if a certain $\bm{k}$-point is invariant under all lattice symmetry operations, nonsymmorphic elements applied to the planewave component may require an additional bare translation as a symmetry element to close the group (see for example p.~56 of Ref.~[\onlinecite{Yu_B10}]). Importantly, the character parity of this translation operating on the planewave component determines whether or not an irreducible representation is indeed physically valid.

\section{\texorpdfstring{$Y$}{Y}-point \label{sec:Ypoint}}
\subsection{Hamiltonian}

The $Y$-point at $(y,z)=(\pm\frac{\pi}{a_y}, 0)$ is left invariant by all point group operations. However, an additional translation is required to close the group of the wavevector. We can see this by considering the noncommutative properties of two symmetry elements acting on a vector in the plane: $\tau C_{2z}R_y(y,z)=(y+\frac{a_y}{2},z+\frac{a_z}{2})$ but $R_y\tau C_{2z}(y,z)=(y-\frac{a_y}{2},z+\frac{a_z}{2})$.

A new operator is required to connect these resulting vectors, $Q_y$ [translation by $(y,z)=(a_y,0)]$. Note that in addition to the $Y$-point itself, the same element is also needed for closure at any $k$-point along the $H$ axis [$Y-T$, see Fig.~\ref{fig:lattice}(b)]. In addition to forming larger classes, i.e. $\mathbb{C}_2\rightarrow \{ \tau C_{2z}, Q_y \tau C_{2z}\}$ (and similarly for $\mathbb{C}_3$ and $\mathbb{C}_4$), $Q_y$ forms an additional new class, $\mathbb{C}_5$. The full character table for the group at $Y$ is shown in Table \ref{tab:Y}. 

\begingroup
\squeezetable
\begin{table} [h!]
\caption{Character table at $Y$-point: $G^4_8$ (see Ref.~[\onlinecite{BC_B10}], p.~ 228, where the IRs are denoted $R_{1-5}$ and classes $C_2=\mathbb{C}_5, C_3=\mathbb{C}_3, C_4=\mathbb{C}_4, C_5=\mathbb{C}_2$). Strikeout indicates terms forbidden by time-reversal symmetry. Only $Y_1$ is physical (odd parity with respect to $\mathbb{C}_5$).\label{tab:Y} }
\renewcommand{\arraystretch}{.6}
\begin{tabular}{c|ccccc|c|c|c|c}
\hline \hline
IR &$\mathbb{C}_1$&$\!2\mathbb{C}_2$&\!2$\mathbb{C}_3$&\!2$\mathbb{C}_4$&$\!\mathbb{C}_5$& basis & invariants & \multicolumn{2}{|c}{$Y_1$ matrices}\\ \hline
$M_1(R_1)$ & 1& 1 &1 &1 & 1& 1, $z$  &$k_y^2$,$k_z^2$, $\cancel{k_z}$, $k_y\sigma_{x}$ & $\rho_0$ &$\varrho_0$\\
$M_2(R_4)$  &1 &1 &-1 &-1 & 1&  $xy, xyz$ &  $\cancel{\sigma_z}$, $k_y\sigma_y$, $k_z\sigma_z$& $\rho_x$ &$\varrho_y$\\
$M_3(R_2)$ & 1& -1 &1 &-1 & 1 &  $y$, $yz$ &  $k_y$, $\sigma_{x}$, $\cancel{k_z\sigma_{x}}$ & $\rho_y$&$\varrho_z$\\
$M_4(R_3)$  & 1& -1 &-1 &1 & 1& $x$, $xz$ & $\cancel{\sigma_{y}}$, $k_y\sigma_z$, $k_z\sigma_{y}$ & $\rho_z$ & $\varrho_x$\\
\hline
$Y_1(R_5)$  & 2& 0 &0 &0 & -2&  & &$\begin{matrix}\{ \cos,\\ \sin\}\end{matrix}$ & $ e^{\pm i\frac{\pi y}{a_y}}$ \\%
  \hline
\end{tabular}
\end{table}
\endgroup

At $Y$, the planewave part of the Bloch wavefunction $e^{i \frac{\pi}{a_y} y}$ undergoes a sign change upon translation by a lattice vector under the symmetry operation $Q_y=(a_y,0)$.  Thus, all valid physical IRs must have odd character in class $\mathbb{C}_5$; only the two-dimensional $Y_1$ meets this requirement. This situation is similar to the double orbital degeneracy of all states at the $X$-point in diamond structure crystals such as Si and Ge, due to nonsymmorphic elements of the diamond symmetry group which cause all singly-degenerate IRs to be unphysical\cite{Yu_B10, Li_PRL11}. In analogy to the IR nomenclature convention in that case, we label the remaining four unphysical IRs at the $Y$-point by $M_{1-4}$; see Table \ref{tab:Y}.

According to the matrix element theorem, all nonzero matrix elements in the effective Hamiltonian are due to interaction terms of symmetry $Y_1 \otimes Y_1 =  M_1 \oplus M_2 \oplus M_3 \oplus M_4$. To proceed using the method of invariants\cite{BP_B74}, we choose a particular set of $Y_1$ basis functions $\{ \cos \frac{\pi y}{a_y}, \sin \frac{\pi y}{a_y}\}$, and generate transformation matrices $D_i$ corresponding to the symmetry operations in classes $\mathbb{C}_{1-5}$.  We find $D_{1-5}=\rho_0,\rho_x,-i\rho_y,\rho_z,-\rho_0$, respectively. Subscripts of these 2$\times$2 matrix operators for orbitals and all $\sigma$ (for spin) and $\tau$ (for remote band coupling) correspond to the usual Pauli matrices, with subscript $0$ indicating the identity. Using these matrix operators, we then invoke the unitary transformation $D_i^{-1} \aleph D_i=\xi_i \aleph$, finding that when $\aleph=\rho_{0,x,y,z}$, the resulting $\xi_i$ are commensurate with distinct IR characters in Table \ref{tab:Y}. We can then assign  $\rho_0 \rightarrow M_1 $, $\rho_x\rightarrow M_2 $,  $\rho_y\rightarrow M_3 $,  $\rho_z\rightarrow M_4 $.

Time-reversal symmetry (TRS) further restricts the presence of some of these spatial-symmetry-allowed terms, i.e. they do not commute with the time-reversal operator $\hat{\Theta}=\hat{K}(\rho_0\otimes \sigma_{y})$ (and $k\rightarrow -k$ due to the planewave part of the wavefunction). We find that Hamiltonian terms proportional to $k_z\rho_0\otimes \sigma_0$, $\rho_x\otimes \sigma_z$, $k_z \rho_y \otimes \sigma_{x}$, and $\rho_z\otimes \sigma_{y}$ are all forbidden. The spin-dependent $4\times 4$ Hamiltonian then has the form
\begin{align}
\mathcal{H}_Y=&A_1 k_y^2\rho_0\otimes\sigma_0+ B k_y \rho_y\otimes \sigma_0 +C \rho_y\otimes\sigma_{x} \notag\\
&+F_1 k_y\rho_z\otimes\sigma_z+F_2 k_y\rho_0\otimes\sigma_{x}+ G_1 k_y\rho_x\otimes\sigma_{y} \nonumber\\
&+A_2 k_z^2\rho_0\otimes\sigma_0 +F_3k_z\rho_z\otimes\sigma_{y}+G_2 k_z\rho_x\otimes\sigma_{z}.
\label{eq:HY}
\end{align}
The first two lines include all $k_y$-related terms together with the only $k$-independent spin-orbit perturbation proportional to constant $C$, while the third line accounts for all $k_z$-related terms. The $F_{1-3}$ terms are lowest-order $k$-dependent spin-orbit perturbations and likely have the same amplitude, because spin-orbit coupling originates from the inner core region where the potential is highly isotropic. The $G_{1-2}$ terms are from higher-order spin-orbit coupling due to the folding of off-diagonal blocks from remote bands \cite{Lowdin_JCP51}; again, their amplitudes are likely to be similar.

The spin-independent Hamiltonian from the three $\sigma_0$ related terms of Eq. \ref{eq:HY} can be diagonalized analytically to find two doubly-degenerate eigenvalues $E_\pm=A_1 k_y^2+A_2k_z^2\pm Bk_y$ corresponding to the eigenvectors of $\rho_y\otimes \sigma_0$.
In other words, the orbital wavefunctions along the $\Delta$-axis ($Y-\Gamma$) are equal superpositions of the basis functions $\ket{\pm}=Y_1^{(1)}\mp i Y_1^{(2)}$. Note that, using our specific basis functions $\{ \cos \frac{\pi y}{a_y}, \sin \frac{\pi y}{a_y}\}$, $\ket{\pm}=e^{\mp i\frac{\pi y}{a_y}}$. While these complex exponential functions are certainly consistent with the spatial symmetry operations, they are not TR invariant since they transform into each other under the complex conjugation operator $\hat{K}$. The time reversal operator $\hat{\Theta}$ in this basis must be modified from its usual form, complicating the selection of TR-allowed invariants. To highlight this issue, we label the invariant matrices for this natural eigenbasis in Table \ref{tab:Y} as $\varrho$, indicating that while they act in the same orbital space as the $\rho$ matrices, they have unusual TRS properties. We will again see the relevance of basis function time-reversal transformation properties in Sec.~\ref{sec:Zpoint} when discussing states near the $Z$-point.

We can use the $\ket{\pm}$ wavefunctions to evaluate the effect of spin-orbit interaction via perturbation theory. 
In the $\varrho\otimes\sigma:\{+\uparrow, +\downarrow, -\uparrow, -\downarrow \}$ unperturbed band basis, all six SOC-related terms of Eq.~(\ref{eq:HY}) are 
\begin{align}
\begin{bmatrix}
0&C&0&0\\
C&0&0&0\\
0&0&0&-C\\
0&0&-C&0
\end{bmatrix}
+
\begin{bmatrix}
0&F_2&F_1&-G_1\\
F_2&0&G_1 &-F_1\\
F_1&G_1&0&F_2\\
-G_1&-F_1&F_2&0
\end{bmatrix}k_y\notag\\
+
\begin{bmatrix}
0&0&-iG_2&-iF_3\\
0&0&iF_3&iG_2\\
iG_2&-iF_3&0&0\\
iF_3&-iG_2&0&0
\end{bmatrix}k_z.
\label{eq:Matrix_Y}
\end{align}
These matrix elements can now be used to analyze the effect of SO perturbation along lines of symmetry. Of particular importance is the case when $k_z=0$ along the $\Delta$-axis ($Y-\Gamma$), since this corresponds to the conduction band minimum ($k_y=k_{y0}\!\sim\!20\%\frac{\pi}{a_y}$) in many `four-six-enes'. Because both the orbital splitting and SOC matrix elements are proportional to $k_y$ along this path, the lowest-order wavefunctions are $k$-independent, e.g. the lowest conduction band is 
\begin{align}
\ket{\psi_c}:\ket{-,x}-\frac{F_1-G_1}{2B}\ket{+,\bar{x}},
\label{eq:Yptwavefnc}
\end{align}
while its spin-split partner, the second lowest conduction band, is
\begin{align}
\ket{\psi_c'}:\ket{-,\bar{x}}-\frac{F_1+G_1}{2B}\ket{+,x}.
\label{eq:Yptwavefncp}
\end{align}
Here, $\ket{x} = \frac{1}{\sqrt{2}}\left(\ket{\uparrow}+\ket{\downarrow}\right)$ and $\ket{\bar{x}} =\frac{1}{\sqrt{2}}\left( \ket{\uparrow}-\ket{\downarrow}\right)$.
In spin space, they are spin-eigenfunctions of $\sigma_{x}$ (out of plane) and slightly mixed since both $|F_1|$ and $|G_1|\ll~\!|B|$. Eq.~(\ref{eq:Matrix_Y}) also shows that around the band extrema, the lowest order Dresselhaus field\cite{Dresselhaus_PR55} follows the form $(C-F_2 k_{y0})\sigma_x$, pointing out of plane.

\begin{figure}
\includegraphics[width=7.5cm]{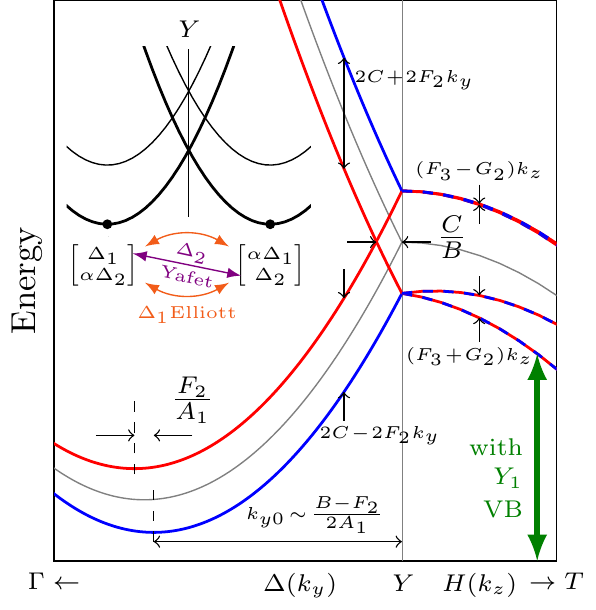}
\caption{(Color online) Bands close to the $Y$-point using Eqs.~(\ref{eq:EDelta}) and (\ref{eq:EH}) with conduction band parameters from Table~\ref{tab:Yparams}. Blue(red) indicates dominant spin orientation up(down) along the out-of-plane $x$ direction; along the $H$-axis  ($Y-T$), spin and orbital degrees of freedom are maximally mixed. Green arrow indicates upward repulsion from interband $\bm k \cdot \bm {\hat{p}}$ coupling with states of lower energy. Gray shows bands before SOI is taken into account. Compare to DFT results in Fig.~\ref{fig:DFT}(a). Inset: Intervalley spin-flip scattering mechanisms and $\Delta$-axis phonon selection rules connecting $x$-axis spinor components, with spin-mixing coefficient $|\alpha|=|\frac{F_1-G_1}{2B}|$.\label{fig:Ybands}}
\end{figure}

The lowest-order eigenvalues along the $\Delta$-axis $(k_z=0)$ are
\begin{align}
E_\Delta=&A_1k_y^2\pm C (+/-)\sqrt{(B\pm F_2)^2+(F_1\pm G_1)^2} k_y \notag\\
\approx &A_1k_y^2 (+/-) \left[B\pm F_2\right]k_y\pm C,
\label{eq:EDelta}
\end{align}
as shown in Fig.~\ref{fig:Ybands}. Here the choice $(+/-)$ selects the upper or lower band of a spin-split pair, whereas the choice $\pm$ determines the majority orbital components of a band. The absolute CBM is thus at $k_{y0}=\frac{B-F_2}{2A_1}$ away from $Y$ toward $\Gamma$, and the shift to upper CB minimum is $\Delta k=F_2/A_1$. In the region of the local conduction band minimum where $k_{y0}\gg k_z$, the wavefunction acquires a higher order spin mixing term $\pm i\frac{(F_3 \pm G_2)k_z}{2Bk_{y0}}$, in addition to the lowest-order mixing given in Eqs.~(\ref{eq:Yptwavefnc}) and (\ref{eq:Yptwavefncp}). 

The lower $\ket{+}$ and upper $\ket{-}$ bands accidentally cross at $C/B$ away from $Y$. It is worth noting that there is no avoided crossing here because, as Table~\ref{tab:Y} shows, there are no spatial symmetry-allowed interactions that couple the $\rho_y$ orbital eigenstates while simultaneously commuting with $\sigma_x$; these interactions coupling majority wavefunction components $\ket{+,\bar{x}}$ and $\ket{-,\bar{x}}$ would take the form  $\rho_{x,z}\otimes \sigma_{x,0}$, but are forbidden by the presence of $\tau R_x$. Neither are there terms that couple majority to minority ($F_1$ and $G_1$ spin-orbit induced) components with identical orbitals but opposite spin, which would be of the form $\rho_{0,y}\otimes \sigma_{y,z}$.  

Now we turn our attention to the situation when $k_y=0$ along the $H$-axis ($Y-T$). The term $F_3k_z\rho_z\otimes\sigma_{y}$ in Eq. \ref{eq:HY} only couples between bands of the same doubly-degenerate subspace, causing an orbital mixing and weak splitting. The two pairs of eigenstates are
\begin{align}
\ket{\psi_l}: \frac{1}{\sqrt{2}}\left(\ket{-,x}\pm i\ket{+,\bar{x}}\right), \ket{\psi_u}: 
\frac{1}{\sqrt{2}}\left(\ket{+,x}\pm i\ket{-,\bar{x}}\right),\label{eq:Y_H_wf}
\end{align}
and the lowest-order eigenvalues along the $H$-axis are 
\begin{align}
E_H=
\left\{\begin{array}{cc}
A_2k_z^2-C\mp(F_3+G_2)k_z,& \mbox{[lower]}\\
A_2k_z^2+C\mp(F_3-G_2)k_z. & \mbox{[upper]}
\end{array}\right.
\label{eq:EH}
\end{align}
The $\mp$ here corresponds to the $\pm$ in the states of Eq. \ref{eq:Y_H_wf}. Notice that the orbital splitting for the upper and lower pairs can be inequivalent, due to constructive and destructive interference between the $F_3$ and $G_2$ parameters.

\begin{table}
\caption{$Y$-point Hamiltonian parameters for SnS.\label{tab:Y_parameters} }
\renewcommand{\arraystretch}{.6}
\begin{tabular}{c|c|c|c|c|c|c}
 & $A_1$[eV\AA$^2$] & $A_2$[eV\AA$^2$] & $B$[eV\AA]& $C$[meV]& $F$[eV\AA] & $G$[eV\AA]\\
\hline
CB & 15 & -8.6 & 4.0& 53 & 0.15 & 0.14\\
VB & -7.4 & 1.0 & 2.1 & 31 & 0.14 & 0.11\\
\hline
\end{tabular}
\label{tab:Yparams}
\end{table}

The fitting of the DFT calculation in Fig. \ref{fig:DFT}(a) determines the numerical values for the parameters $A_{1,2}, B, C, F=F_{1,2,3},$ and $G=G_{1,2}$ given in Table~\ref{tab:Y}. Fig.~\ref{fig:Ybands} shows the resulting bands described by Eqs.~\ref{eq:EDelta}~and~\ref{eq:EH} along the $\Delta-$ and $H-$axes, with colors red/blue indicating the spin-split nature along the former. It should be emphasized that the full wavevector dependence of the four bands, 
\begin{align}
&E_Y(k_y,k_z)=A_1k_y^2+A_2k_z^2 \pm C \notag\\
&(-/+)\sqrt{[(B\pm F_2)^2+(F_1 \pm G_1)^2]k_y^2+(F_3 \mp G_2)^2 k_z^2}. \notag
\end{align}
is obtained from exact diagonalization of the lowest-order approximate Hamiltonian, expected to be most valid close to the $Y$-point and along these high symmetry axes. 

\subsection {Interband interaction}

The discussion above only includes interactions between the two $Y_1$ states that are orbitally-degenerate at the $Y$-point. Coupling to nearby bands, all of which are also of $Y_1$ symmetry, are via off-diagonal blocks of the matrix Hamiltonian. In addition to the terms appearing on the main diagonal blocks, TRS allows otherwise-forbidden invariant components here. We can formally include them by projecting into the larger Hilbert space via direct product with the $\tau_x$ and $\tau_y$ operators, respectively. The effect of SOC due to folding these interactions into the 4$\times$4 Hamiltonian has already been discussed, giving rise to the terms proportional to $G_1$ and $G_2$. We emphasize that, unlike the case of inversion-symmetric lattices, the most obvious signature of SOC near $Y$ (spin splitting) is independent of coupling to remote bands, similar to the case of states near the $K$- and $K'$- points in transition-metal dichalcogenides.\cite{Mak_NatPhot15}

The most important interband terms that give rise to band dispersion are those that arise from the spin-independent $\bm{k}\cdot \bm{\hat{p}}$ perturbation. In addition to a coupling of the form $\tau_x\otimes k_y\rho_y\otimes\sigma_0$,  there is also a term (previously forbidden by TRS on the diagonal block) proportional to $\tau_y\otimes k_z\rho_0\otimes\sigma_0$. 
Because $A_1$ and $A_2$ are of opposite sign, the contribution from coupling to the valence bands via the latter term is necessary to fully reproduce the absolute global conduction band minimum (rather than a saddle point) at ($k_{y0},0$) and repel the bands upward along $k_z$ to preserve an open bandgap.

\subsection{Optical selection rules}

The interband coupling mentioned above is also useful in the context of analyzing optical selection rules at the $Y$-point and the conduction valleys close to it. Interaction of electromagnetic excitation is provided by a dipole perturbation that transforms as a polar vector, just like the $\bm k \cdot \hat{\bm p}$ terms. Among the three components of linear polarization,  the one directed out of the plane ($\hat{x}$) is associated with IR $M_4$, which behaves as the off-diagonal $\varrho_x$ that couples different $|\pm\rangle$ orbital basis functions. By checking the numerical wavefunctions of band-edge states at the $Y-\Gamma$ valleys, we confirm that their orbital constituents are identical, indicating that this $\hat{x}$-polarization is forbidden, whereas the in-plane $\hat{y}$ and $\hat{z}$ are allowed (associated with IR $M_3$ and $M_1$, respectively). This can be contrasted with luminescence in the $D_{3h}$ group-III metal monochalcogenides, whose band extrema are coupled to lowest order only by out-of-plane polarized emission\cite{Li_PRB15}.

The amplitude of the $z$-polarized dipole depends on the parity components of the interband states with respect to $\hat{z}$, which are not restricted by any symmetry of the $Y$-point. By further examining the numerical wavefunctions from DFT, we find that the two band edge states near $Y$ are both dominated by planewave components even in $\hat{z}$, so that the matrix element of the odd-parity $\hat{p}_z$ operator between these components is strongly suppressed. We can therefore straightforwardly illuminate the reason behind a dramatic difference of optical transition rates in the $Y-\Gamma$ valleys for the two orthogonal linear polarizations, a result originally obtained by empirical observation of DFT calculations and reported without explanation\cite{Rodin_PRB16}. Our further inspection of spin-dependent wavefunctions reveals that states at the $Y$-valley extrema share the same majority spin after spin-splitting, which guarantees direct optical transition spectra starting from the absolute band edge.


\subsection{Momentum and spin relaxation}

Conduction-band-minimum states lie on the $\Delta$-axis, whose symmetry group consists only of vertical reflection $R_x$ besides the identity. Thus, electron states are of either $\Delta_1$ (even) or $\Delta_2$ (odd) symmetry with respect to $R_x$, and correspond directly to the two $Y$-point basis functions $\ket{+}$ and $\ket{-}$. 

{\em Intravalley--} Zone-center phonons with vanishing momentum that couple these states to themselves, and thus drive intravalley scattering, must be even with respect to $R_x$ and hence are of $\Gamma_1$ ($z$-polarized acoustic) or $\Gamma_3$ ($y$-polarized acoustic or $yz$ optical) symmetry, using compatibility $M_{1-4}\rightarrow\Gamma_{1-4}$ and characters in Table \ref{tab:Y}. Both of these phonons can be efficiently suppressed at low temperatures because, unlike the $x$-polarized flexural mode that has quadratic dispersion\cite{Song_PRL13}, they have a vanishing density of states (acoustic $y$, $z$) or an energy gap (optical $yz$).

{\em Intervalley--} Since $\Delta_1$ and $\Delta_2$ bands cross at the $Y$-point, the dominant wavefunction component for states in the two degenerate valleys are of opposite parity with respect to $R_x$. Phonon-induced transitions then have selection rules reminiscent of spin-conserving \textit{g}-process intervalley scattering in Si, where $\Delta_1$ and $\Delta_2'$ bands cross at the $X$-point so that a $\Delta_2'$ phonon is required\cite{Streitwolf_PSS70}.

In four-six-enes, however, the absence of spatial inversion symmetry breaks spin degeneracy along the $\Delta$-axis and leaves two degenerate valley minima of opposite spin, eliminating the spin-conserving intervalley scattering channel. The remaining allowed spin-flip transitions from the conduction-band minimum state $|\psi_c\rangle$ (see Eq. \ref{eq:Yptwavefnc}) into its time-reversed counterpart close to the absolute conduction band edge with wavefunction 
\begin{align}
\ket{\bar{\psi}_c}: \ket{+,\bar{x}}-\frac{F_1-G_1}{2B}\ket{-,x}
\label{eq:psi_c_Yn}
\end{align}
requires $\Delta_1$-phonon ($y$, $z$, or $yz$-polarized) coupling between majority and minority (spin-orbit mixed) components in the bare Elliot mechanism\cite{Elliott_PR54}.  Since $F$ and $G$ coefficients are nearly the same magnitude in SnS [as can be seen visually in Fig. \ref{fig:DFT}(a) from the weak splitting of upper states along $k_z$], the spin mixing amplitude proportional to $F_1-G_1$ is quite small (approximately 0.1\% from our DFT result) and spin relaxation through this otherwise dominant mechanism is relatively slow. We note that this determination of the matrix element magnitude would be impossible if only analyzing in the $\Delta_{1,2}$ basis without incorporating the influence of the nearby $Y$-point.

Another contribution to intervalley spin relaxation from the Yafet mechanism  that couples majority components with opposite spin\cite{Yafet_SSP63} requires phonons of $\Delta_2$ symmetry, which includes the out-of-plane $x$-polarized and $xy$ or $xz$ modes. It is of the same order of magnitude as the Elliott term; both are indicated in an inset to Fig. \ref{fig:Ybands}, where we use $x$-axis spinor notation to denote the wavefunctions and define $|\alpha|=|\frac{F_1-G_1}{2B}|$.

\section{\texorpdfstring{$Z$}{Z}-point \label{sec:Zpoint}}

Referring to Fig. \ref{fig:DFT}(b), we can see that the valence band maximum is near the $Z$-point. We now modify the approach used above for $Y$-point to analyze these states. 

\subsection{Hamiltonian}
At the $Z$-point, $(y,z)=(0,\pm\frac{\pi}{a_z})$, the translation operator necessary to close the group creates product operations with the point-group symmetries that form distinct new classes, with a total of eight; see Table \ref{tab:Z}. The required odd parity under $Q_z=(0,a_z)$ (again caused by the planewave part of the wavefunction) constrains the physical IRs to $Z_1,Z_{1'},Z_2,Z_{2'}$.

Note that the pairs $Z_1\oplus Z_{1'}$ and $Z_2 \oplus Z_{2'}$ form degenerate ``time-reversal conjugate representations" \cite{Dresselhaus_B08}, i.e. they transform into each other upon time reversal (complex conjugation in single-group). Within a doubly degenerate subspace, 
$Z_1^* \otimes Z_{1'}=Z_2^* \otimes Z_{2'}=M'_4$,
and coupling to bands of the same IR always gives $M'_1$. We therefore only expect terms proportional to $k_{y,z}^2$ and $k_z$ (both belong to $M'_1$) to appear in the spin-independent Hamiltonian.

However, because spatial and time-reversal symmetries are conflated, it is intuitive to define a time-reversal-invariant basis $Z_+=\frac{1}{\sqrt{2}}(Z_1+Z_{1'}), Z_-=\frac{i}{\sqrt{2}}(Z_1-Z_{1'})$ so that $\hat{\Theta}=\hat{K}\rho_0$ as usual. 
Transforming into this basis, we find that the spatial symmetry operations involving partial translation ($\tau C_{2z}$ and $\tau R_y$) have matrix representations $D_i\propto\rho_y$, and all other operations with real-valued characters are $\pm \rho_0$.

Inspecting the matrices satisfying the transformation $D_i^{-1} \aleph D_i=\xi_i \aleph$ leads us to assign $\aleph=\rho_0,\rho_y\rightarrow M_1'$ and $\rho_x,\rho_z\rightarrow M_4'$ in Table~\ref{tab:Z}. We then find that TRS restricts the spin-independent lowest-order effective Hamiltonian to $(A_1'k_z^2+A_2'k_y^2)\rho_0+B'k_z\rho_y$ in the $\{Z_+,Z_-\}$ basis, with eigenvalues $E_\pm=A_1'k_z^2+A_2'k_y^2\pm B'k_z$ and corresponding eigenvectors of $\rho_y$ [$\ket{\pm}=\frac{1}{\sqrt{2}}(Z_+\pm iZ_-)=Z_{1',1}$]. As a result, the degeneracy is broken away from the $Z$-point, with the only exception being the Brillouin zone edge $B$-axis ($Z-T$, $k_z=0$), where the TRS-induced double degeneracy is preserved.

\begingroup
\squeezetable
\begin{table}
\caption{$Z$-point character table: $G^2_8$ (see Ref.~[\onlinecite{BC_B10}], p.~227. Characters of $\mathbb{C}_4$, $\mathbb{C}_7$ and $\mathbb{C}_8$ are not included). $M_{1-4}'$ share the same basis functions accordingly with $M_{1-4}$ in Table~\ref{tab:Y}. Backward cancel indicates combination with the first invariant matrix listed in the RHS column is forbidden by TRS; forward indicates the second. \label{tab:Z} }
\renewcommand{\arraystretch}{.5}
\begin{tabular}{c|ccccc|c|c|c}
&$\mathbb{C}_1$ & $\mathbb{C}_2$ & $\mathbb{C}_6$ & $\mathbb{C}_5$ & $\mathbb{C}_3$ &\multirow{2}{*}{invariants}&$\{Z_+,$ &$\{Z_{1(2)},$ \\
& $E$ & $\tau C_{2z}$ & $\tau R_x$ & $R_y$ & $Q_z$ & & $Z_-$\}& $Z_{1'(2')}$\}\\
\hline 
$M'_1(R_1)$ &     1& 1& 1& 1& 1 & $\cancel{k_y^2},\cancel{k_z^2}, \bcancel{k_z},\cancel{k_y\sigma_{x}}$&$\rho_0,\rho_y$& $\varrho_0,\varrho_z$\\ 
$M'_2(R_5)$  &    1 &1 &-1& -1 & 1 &$\sigma_z, k_y\sigma_y, k_z\sigma_z$&&\\
$M'_3(R_7)$  &    1 &-1 & 1 &-1 & 1 &$k_y,\sigma_{x},k_z\sigma_{x}$&&\\
$M'_4(R_3)$   &   1& -1 &-1& 1 & 1 &$\xcancel{\sigma_{y}},k_y\sigma_z, k_z\sigma_{y}$ &$\rho_x,\rho_z$&$\varrho_x,\varrho_y$\\
\hline
$Z_1(R_2)$ &   1& i& i& 1 & -1& & & \\ 
$Z_{1'}(R_4)$ &   1& -i &-i& 1 & -1& & &\\
$Z_2(R_6)$ &   1 &i &-i& -1 & -1& & &\\
$Z_{2'}(R_8)$ &   1 &-i& i &-1 & -1& & &\\

\hline
\end{tabular}
\end{table}
\endgroup

Inclusion of spin and SO coupling is straightforward. The combination of spatial symmetry and time-reversal symmetry ($\hat{\Theta}=\hat{K}\rho_0\otimes\sigma_y$) excludes all $k$-independent SOC as shown in Table~\ref{tab:Z}; as a result, the degeneracy of both $Z_1\oplus Z_{1'}$ and $Z_2\oplus Z_{2'}$ bands is doubled to fourfold at the $Z$-point, even though the only physically-allowed IR is two dimensional in the $Z$-point double group $G_{16}^8$ (see Ref.~[\onlinecite{BC_B10}], p.~233). 

The only TRS-allowed SO term of $M_1'$ symmetry is $k_y\rho_0\otimes \sigma_x$, whose strength we parameterize in the Hamiltonian by the constant $F_1'$. There are four TRS-allowed SO terms of $M_4'$ symmetry: $k_y\{ \rho_x,\rho_z\}\otimes \sigma_z$, and $k_z \{ \rho_x,\rho_z\}\otimes  \sigma_{y}$. However, only two parameters $F'_2$ and $F'_3$ are required to obtain the eigenvalue spectrum; the coefficients weighting each $\rho_x,\rho_z$ term are merely {\em{orbital}} basis-dependent quantities and hence are only necessary to determine the exact details of the wavefunctions beyond symmetry. The full Hamiltonian expressed in the $\{Z_+,Z_-\}$ basis is then 
\begin{align}
\mathcal{H}_Z=&A_1'k_z^2\rho_0\otimes \sigma_0+A_2'k_y^2\rho_0\otimes \sigma_0+B'k_z\rho_y\otimes \sigma_0\notag\\
& +F_1'k_y\rho_0\otimes\sigma_{x} +F_2'k_y\rho_{z,x}\otimes\sigma_z +F_3'k_z\rho_{z,x}\otimes\sigma_{y}.\notag
\end{align}
In the natural (but not TR invariant) $\{Z_1,Z_{1'}\}$ eigenbasis, it becomes
\begin{align}
\mathcal{H}_Z=&A_1'k_{z}^2\varrho_0\otimes \sigma_0+A_2'k_{y}^2\varrho_0\otimes \sigma_0+B'k_z\varrho_z\otimes \sigma_0\notag\\
&+ F_1'k_y\varrho_0\otimes\sigma_{x} +F_2'k_y\varrho_{y,x}\otimes\sigma_z +F_3'k_z\varrho_{y,x}\otimes\sigma_{y}.\label{eq:HZ2}
\end{align} 
Note that we can repeat the above analysis with $Z_2\oplus Z_{2'}$ and obtain the same result. 


\begin{figure}
\centering
\includegraphics[width=7.5cm]{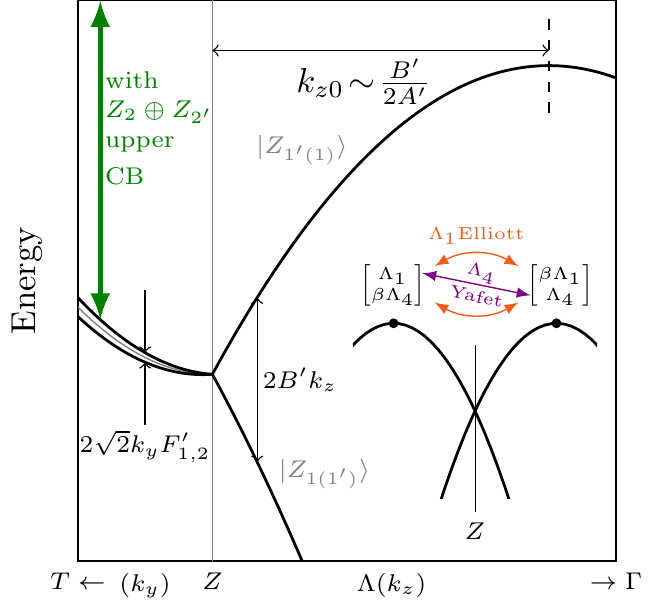}
\caption{(Color online) Bands close to the $Z$-point using Eqs.~(\ref{eq:EB}) and (\ref{eq:ELambda}) with SnS valence band parameters from Table \ref{tab:Zparams}. Along the $\Lambda$-axis ($Z-\Gamma$), bands remain spin-degenerate.  Gray shows bands before SOI is taken into account and is visible along the $B$ axis ($Z-T$). Green arrow indicates downward repulsion from interband $\bm k \cdot \bm {\hat{p}}$ coupling with states of higher energy. Compare to DFT results in Fig.~\ref{fig:DFT}(b). Inset: Intervalley spin-flip scattering mechanisms and $\Lambda$-axis phonon selection rules connecting $x$-axis spinor components. \label{fig:Zbands}}
 \vspace{-0.2in} 
\end{figure}

\begin{table}
\caption{$Z$-point Hamiltonian parameters for SnS.\label{tab:Z_parameters} }
\renewcommand{\arraystretch}{.6}
\begin{tabular}{c|c|c|c|c}
 & $A_1'$ [eV\AA$^2$] & $A_2'$ [eV\AA$^2$] & $B'$ [eV\AA]& $F'$[eV\AA]\\
\hline
CB & 13.8 & -8.9 & 4.0 & $ 0.40 $\\
VB & -11.0 & 15.0 & -3.3 & $ -0.09 $\\
\hline
\end{tabular}
\label{tab:Zparams}
\end{table}

The spectrum along the Brillouin zone edge $B$-axis ($Z-T$) is given by exact diagonalization of Eq.~(\ref{eq:HZ2})
\begin{align}
E_B=A_2'k_y^2(+/-) k_y\sqrt{F_1'^2+F_2'^2},
\label{eq:EB}
\end{align}
where $(+/-)$ selects between the two spin states. The corresponding wavefunctions are both orbitally and spin mixed, with amplitudes depending on the relative strength of $F'_1$ and $F'_2$.

Along the $\Lambda$-axis ($Z-\Gamma$, $k_y =0$), spin degeneracy is preserved; the last term in Eq.~(\ref{eq:HZ2}) clearly only couples between states of opposite spin in different orbital bands. Thus, the degenerate wavefunctions can be expressed as spin- and orbitally-mixed, i.e.
\begin{align}
\ket{\psi_v,\Uparrow}=\ket{Z_{1(1')}, x}- \frac{F_3'}{2B'}\ket{Z_{1'(1)}, \bar{x}},
\label{eq:Z_wf_up}\\
\ket{\psi_v,\Downarrow}=\ket{Z_{1(1')} ,\bar{x}}+ \frac{F_3'}{2B'}\ket{Z_{1'(1)}, x},
\label{eq:Z_wf_dn}
 \end{align}
where we have indicated that the orbital functions swap roles between the two equivalent valleys. The energy spectrum is 
\begin{align}
E_\Lambda=A_1'k_z^2\pm k_z\sqrt{B'^2 +F_3'^2},\notag\\
\approx A_1'k_z^2\pm (B' +\frac{F_3'^2}{2B'})k_z
\label{eq:ELambda}
\end{align}
where $\pm$ selects between orbital states. Away from the $\Lambda$-axis in the $k_y$ direction, the fourth term in Eq.~(\ref{eq:HZ2}) causes a lowest order linear Dresselhaus splitting of the two states in Eqs. (\ref{eq:Z_wf_up}) and (\ref{eq:Z_wf_dn}), similar in form to the case of [110] zincblende quantum wells\cite{Ohno_PRL99, Dohrmann_PRL04, Couto_PRL07, Muller_PRL08}.  Our analytic dispersion near the valence band $Z$-point can then be plotted using Eqs. (\ref{eq:EB}) and (\ref{eq:ELambda}), and the parameters in Table~\ref{tab:Z} taken from fitting the DFT results in Fig.~\ref{fig:DFT}(b).

\subsection {Interband interaction}

Coupling to other bands of opposite single-group symmetry gives different invariant matrices $\rho_0,\rho_y\rightarrow M_2'$ and $\rho_x, \rho_z\rightarrow M_3'$. Once again, TRS is no longer an absolute constraint on these coupling matrices as long as the full Hamiltonian remains TR invariant. We thus have many symmetry-allowed terms in the $(Z_1\oplus Z_{1'})\leftrightarrow (Z_2\oplus Z_{2'})$ block:
\begin{align}
\tau_x\otimes\{ \varrho_z\otimes\sigma_z, k_z\varrho_{x,y}\otimes\sigma_{x}\}
\end{align}
and
\begin{align}
\tau_y\otimes\{ \varrho_0\otimes\sigma_z, \varrho_{x,y}\otimes\sigma_{x}, k_y\varrho_{x,y}\otimes \sigma_0   \}.\label{eq:Zintop}
\end{align}
When $k_y=0$, this perturbation only couples each band to two other remote states, all of which are the same spin. The two matrix elements have squared magnitude $\alpha ^2+\delta ^2$ for the same  orbital, and $(\phi\pm\beta k_z)^2 +(\eta \mp \chi k_z)^2$ for the opposite orbital, where $\alpha,\beta,\chi, \delta, \eta,\phi$ are coefficients of the above $k_y$-independent SOC terms, respectively. Although the $\pm$ causes this shift to take on different $k_z$-dependent values for the two orbitals within a given band, they are spin independent and thus twofold degeneracy is maintained along the $\Lambda$-axis ($Z-\Gamma$). Importantly, the last $k_y$-related $\bm{k}\cdot \bm{\hat{p}}$ terms in Eq.~(\ref{eq:Zintop}) induced from $Z_2\oplus Z_{2'}$ upper conduction band states are responsible for the downward repulsion  of the valence band maximum along $k_y$, which opens a bandgap.

For coupling to remote bands of the same symmetry, we have off-diagonal terms $\tau_x\otimes$
\begin{align}
&\{k_{y,z}^2\varrho_0\otimes \sigma_0,  k_z\varrho_z\otimes \sigma_0,  k_y \varrho_0\otimes\sigma_{x},  k_y\varrho_{x,y}\otimes\sigma_z\}, \label{eq:Z_off_block_1}
\end{align}
and $\tau_y\otimes$
\begin{align}
& \{k_{y,z}^2\varrho_z\otimes \sigma_0, k_z\varrho_0\otimes \sigma_0,k_y\varrho_z\otimes\sigma_{x},
\varrho_{x,y}\otimes\sigma_{y} \},\label{eq:Z_off_block_2}
\end{align}
for $(Z_1\oplus Z_{1'})\leftrightarrow (Z_1\oplus Z_{1'})$, or $(Z_2\oplus Z_{2'})\leftrightarrow (Z_2\oplus Z_{2'})$ matrix blocks. A similar analysis as in the previous case leads us to the same result, i.e. no additional spin splitting. This is confirmed by examination of the only physically-allowed IR in the double-group character table for the $\Lambda$-axis ($G^5_8$, see Ref.~[\onlinecite{BC_B10}], p.~228), which is of dimension two. 

By examining the numerical wavefunctions from our DFT calculation, we find that both the lowest conduction bands and the highest valence bands at the $Z$ point belong to the $(Z_1\oplus Z_{1'})$ pair, which is even under $R_y$ and favored by states with relatively low energy (compared with the $Z_2\oplus Z_{2'}$ pair that is odd under $R_y$). At a fixed $k$ point along the $\Lambda$ axis, the conduction and valence valleys belong to the same IR (both $Z_1$ or $Z_{1'}$), which is evident by their identical parity under the operation of $\tau R_x$. We can then see from Eqs. (\ref{eq:Z_off_block_1}) and  (\ref{eq:Z_off_block_2}) that the two $k_z$-dependent terms are those responsible for band-to-band repulsion and the opening of a bandgap along the $\Lambda$-axis ($Z-\Gamma$). 

\subsection{Optical selection rules}
Because both the upper valence and lowest conduction bands pairs at the $Z$-point are of $Z_1\oplus Z_{1'}$ symmetry, we can use the same expressions to gain insight into the optical selection rules for direct excitation from the $Z-\Gamma$ valence valleys. In particular, we see that the $k_y$ component of the $\bm{k}\cdot \bm{\hat{p}}$ perturbation is forbidden by spatial symmetries, so $\hat{y}$-polarized light does not induce lowest order optical transitions requiring nonzero $\hat p_y$ matrix elements. On the other hand, direct gap $\hat{z}$-polarized optical excitation is allowed due to the same symmetry behavior of the $Z-\Gamma$ conduction and valence valleys ($Z_{1(1')}^*\otimes Z_{1(1')}=M_1'$, with $z$ as a basis function).


\subsection{Momentum and spin relaxation}

The $\Lambda$-axis has point-group symmetry identical to that at $\Gamma$. However, since the $Z_{1(1')}$ basis states that form valence band wavefunctions are both even with respect to reflection operator $R_y$, they correspond only to either $\Lambda_1$ (transforms like $1$, $z$) or $\Lambda_4$ (transforms like $x$, $xz$). Of course, spin-orbit interaction injects weak admixtures of opposite spin and spatial symmetry into these otherwise pure states (see Eqs. \ref{eq:Z_wf_up} and \ref{eq:Z_wf_dn}).

{\em Intravalley--} Once again, spin-preserving intravalley scattering within a band of definite $\Lambda_1$ or $\Lambda_4$ symmetry is driven by $z$-polarized $\Gamma_1$ zone-center acoustic phonons. This phonon is thus the primary source of intrinsic hole-transport mobility limitation in the valence band.

Both the out-of-plane $x$-polarized flexural phonon and the $xz$ optical phonon of $\Gamma_4$ symmetry will drive intravalley spin-flips by coupling majority to minority wavefunction components with squared matrix element proportional to $\beta^2$, where $\beta=F'_3/2B'$. Although the optical phonon branch is gapped and can be suppressed by reducing thermal energy temperature, flexural phonons have a constant density of states so are present in high densities even at low temperatures\cite{Song_PRL13}. They are thus the primary source of spin relaxation for states along the $\Lambda$-axis. 

The valence band valley effective mass is nearly isotropic, however, so we must consider states off the primary $\Lambda$-axis. Due to linear Dresselhaus splitting for states in the $k_y$ direction away from the band extremum, we expect the total spin relaxation to be highly anisotropic, providing a means to control the spin lifetime via interplay of the intrinsic Dresselhaus field and a gate-induced Rashba field \cite{Appelbaum_PRA16}. Although in-plane polarized spins with $k_y\neq~\!\!0$ will rapidly precess, \cite{Dyakonov_SPSS1972} the quadratic dispersion and linear spin splitting give rise to a `persistent spin helix'\cite{Schliemann_PRL03, Koralek_Nature09, Bernevig_PRL06} during the process of lateral diffusion. Using our nomenclature for the spin-orbit parameters and assuming an isotropic valence-band effective mass, this helix will have a wavelength $\approx 2\pi\frac{A_1'}{\sqrt{F_1'^2+F_2'^2}}\approx$50~nm in SnS.

{\em Intervalley--} $\Lambda$-axis phonons will drive intervalley scattering across the $Z$-point. Unlike the previous case of intervalley scattering across the $Y$-point, both spin-flip and spin-preserving scattering are allowed to lowest order because of spin degeneracy on the $\Lambda$-axis, similar to the \textit{g}-process in Si.\cite{Li_PRL11,Streitwolf_PSS70} 

Spin-preserving intervalley transitions are induced by phonons with $\Lambda_4$ symmetry, polarized either in the out of plane $x$-direction or of $xz$ symmetry; the same phonons drive the intervalley Yafet spin-flip mechanism because it similarly couples majority wavefunction components. The intervalley Elliott mechanism is instead driven by $\Lambda_1$ phonons with in-plane $z$ polarization. The matrix element is similar to that for intravalley Elliott spin flips with squared magnitude $\beta^2$. Both intervalley spin-flip processes are indicated in an inset to Fig. \ref{fig:Zbands}, where we use $x$-axis spinor notation to denote the band-edge wavefunctions. 

\section{conclusions\label{sec:conclusion}}

We began this paper by interrogating several salient features of the first-principles bandstructure. We then used the full theory of space groups and method of invariants relevant to the $Y$- and $Z$-points to not only uncover the symmetries explaining the origins of these eigenvalue dispersion peculiarities, but also to predict selection rules for wavefunction transitions.  

We explained why the states at the $Y$-point are split by spin-orbit coupling but those at the $Z$-point are not. We identified Hamiltonian terms forbidden by symmetry which preserve accidental band crossings near $Y$, and spin degeneracy near $Z$. We showed that flexural phonons are not the primary source of either intravalley momentum or spin scattering near $Y$, but that they cause efficient spin-flip Elliott scattering near $Z$. We revealed the cause for a weak direct optical dipole along $\hat z$ at the $Y$ valleys and the complete absence of transitions induced by $\hat y$-polarized light in the $Z$ valleys.

Some final comments are intended to guide future research on this material, especially with experimental techniques:

Investigation of spin-related properties in semiconductors often begins with efforts to generate polarized carriers via optical orientation\cite{Dyakonov_SPJETP71}. This approach is fruitful for materials whose electronic structure contains band-edge states where an orbital degeneracy is broken by SOC, as in the valence-band split-off gap in cubic semiconductors. Four-six-enes have a spin-orbit-induced broken degeneracy in the $Y$ valleys, but two symmetry-related effects preclude optical orientation by direct electromagnetic transitions: First, the in-plane $y$ and $z$ directions are inequivalent (which otherwise allows circularly-polarized photoexcitation in the cubic system to generate spin). Second, time-reversal symmetry inverts the spin splitting in the two equivalent $Y$ valleys, with no way to create an imbalance in excitation rates. As is the case in phosphorene\cite{Li_PRB14}, electrical injection from ferromagnetic tunnel contacts are thus required to study spin relaxation effects we have predicted. 

It has been suggested that the valley-dependent sensitivity of interband transitions crossing the bandgap in four-six-enes to linear optical polarization can be used to create a so-called `valleytronics' device in which information is encoded in valley population. However, $f$-type intervalley scattering\cite{Morin_PR57, Song_PRL14} that couples band minima on orthogonal axes will of course interfere with this application. In fact, if it is fast enough, this relaxation can populate the band extrema to create large numbers of crystal momentum-indirect excitons under strong pumping conditions. Because the lifetime of such excitons is typically very long, this material may be a host for creating a degenerate Bose gas of excitons with substantial condensate fraction\cite{Eisenstein_Nature04}. Other indirect-gap few-layer semiconductors, such as strained TMDCs \cite{Horzum_PRB13}, TM-trichalcogenides\cite{Li_Nanoscale15} or  arsenene\cite{Luo_SCPMA15} may also be an option for this research direction, assuming their optical properties are amenable.

\begin{acknowledgments}
We thank Prof. J.D. Sau for illuminating discussions. This work was supported by the Office of Naval Research under contract N000141410317, and the Defense Threat Reduction Agency under contract HDTRA1-13-1-0013.
\end{acknowledgments}

%

\end{document}